\documentclass[runningheads]{llncs}
\usepackage[T1]{fontenc}
\usepackage{lmodern}
\usepackage{amsmath,amssymb,booktabs,tabularx,array}
\usepackage{graphicx}
\usepackage{enumitem}
\usepackage{microtype}
\usepackage[hidelinks]{hyperref}
\setlist{nosep,leftmargin=*}
\newcommand{\code}[1]{\texttt{\small #1}}
\begin{document}
\title{TrustBound: Mechanized Completeness for Payment Epochs Initiated by AI Agents}
\titlerunning{TrustBound: Mechanized Payment-Epoch Completeness}
\author{Sandip Nandi \and Aishwarya Bose \and Apurba Bose \and Soovadeep Bakshi}
\authorrunning{S. Nandi et al.}
\institute{Independent Researchers}
\maketitle

\begin{abstract}
Stopping an autonomous agent does not stop the effects it has already set in
motion. Payment operations requested by an AI agent may remain executable, or
their outcomes unresolved, after the agent has stopped. We study when a trusted
payment service can account for every payment in an epoch and exclude further
payments under earlier authority. We give a Lean model in which operations are
recorded before they receive execution authority and both actions are
serialized with an admission fence. We prove that fencing fixes a journal that
covers every modeled effect and outstanding right; that discharging those
rights prevents further effects; and that truthful outcomes for the recorded
identities then give exact accounting. A provider-contract model, mapped to the
admission model, makes created payment objects inert until registered.
Premise-removal mutations confirm that each key premise is needed. We connect
the models to an executable service: six service histories are checked against
the model in Lean, and bounded schedules, scripted clients and concurrent
workers exercise its reconciliation. The concurrent experiments uncovered a
conflict caused by reordered provider reads, which serializing retrieval with
journal updates, and read tickets for unserialized or cached reads, remove in
all tested cases. The result separates two
obligations that are easily conflated in practice---preventing future effects
and accounting for past ones---and states precisely which guarantees of the
trusted component that mediates an autonomous system's effects establish each.
\keywords{Theorem proving \and Lean \and Runtime verification \and Autonomous agents \and Payments}
\end{abstract}

\section{Introduction}
\label{sec:intro}
A travel business uses software agents to book a flight and a hotel. The hotel
agent requests a \$200 payment through the business's payment service, which
authorizes it and sends it to the payment provider. No reply arrives before the
timeout, so the business pauses the agent for review. Pausing the agent stops
new requests, but it does not recall the request already sent: the hotel
payment may still be queued at the provider and may complete minutes later. Or
it may already have completed, with only the reply lost. Before closing the
trip's books, an operator must answer two different questions. \emph{Can any
further payment still happen under authority granted earlier?} And \emph{which
payments actually happened?}

These questions are routinely conflated. Stopping the agent, revoking an API
key, or rejecting new requests addresses neither: earlier authority can still
execute. Conversely, knowing that nothing more can execute does not reveal the
outcome of a payment whose reply was lost. Card networks now promote
agent-initiated payments with exactly these flight and hotel
scenarios~\cite{mastercard2026,visa2026}, and an agent can make the right
booking decision while leaving a payment unresolved.

The same gap arises wherever an autonomous system acts on the world through a
trusted intermediary: a vehicle, robot or spacecraft whose controller is halted
while commands it issued are still queued or executing downstream, with some
acknowledgements lost. Safe shutdown then needs two separate assurances: that no
further effect can occur under authority already granted, and a complete
account of the effects that did occur. Payments are a demanding instance of
this problem, because effects are executed by an external provider, are
irreversible once terminal, and may be reported late or not at all. We use
payments throughout to keep the model concrete; the admission discipline
depends only on identities, rights, effects and outcomes.

We call the first property \emph{future-effect exclusion} and the combination
of both \emph{exact closure} of a payment epoch. This paper asks: \emph{what
must a trusted payment service guarantee so that exact closure of an epoch can
be established, and how can an implementation be checked against those
guarantees?}

Our answer is an admission discipline with three rules. (1)~\emph{Record before
granting:} every operation's identity enters a durable journal before it
receives authority to produce an effect. (2)~\emph{Fence admission:} recording
and granting are serialized with a fence that stops new admissions, fixing the
journal's population. (3)~\emph{Discharge and report:} once no remaining right
can produce an effect, truthful outcomes for the fixed journal give exact
identity-level accounting. Neither a fence nor drainage alone suffices, and we
exhibit both failures as checked counterexecutions.

\paragraph{Contributions.}
\begin{enumerate}
\item A Lean admission model and closure theorem showing that the three rules
yield a fixed covered population, future-effect stability, and exact accounting
of past effects, with counterexecutions for a fence without drainage and for
an unmediated route (Sect.~\ref{sec:theorem}).
\item A provider-contract model, mapped to the admission model in Lean, that
captures inert payment-object creation, registration before activation, and
irreversible terminality (Sect.~\ref{sec:mapping}).
\item An executable service implementing the discipline, whose recorded
histories are projected into the model and checked in Lean, a form of offline
runtime verification~\cite{leucker2009}
(Sects.~\ref{sec:impl}--\ref{sec:eval}).
\item An evaluation with premise-removal mutations, an evidence-rule
diagnostic, 864 bounded schedules, scripted request-only clients, and 96
concurrent-worker cases, which uncovered and repaired a reconciliation conflict
caused by reordered provider reads (Sect.~\ref{sec:eval}).
\end{enumerate}
Effect closure, commit-time validation and complete mediation are established
principles~\cite{effectbound,cta,salas,saltzer1975}. Our contribution is to make
them precise and checkable for payments, where authority, provider objects and
outcomes have specific structure, and to connect the resulting theorem to an
implementation. All artifacts are public (Sect.~\ref{sec:conclusion}).

\section{System Model and Problem Statement}
\label{sec:model}
\begin{figure}[t]
\centering
\includegraphics[width=\linewidth]{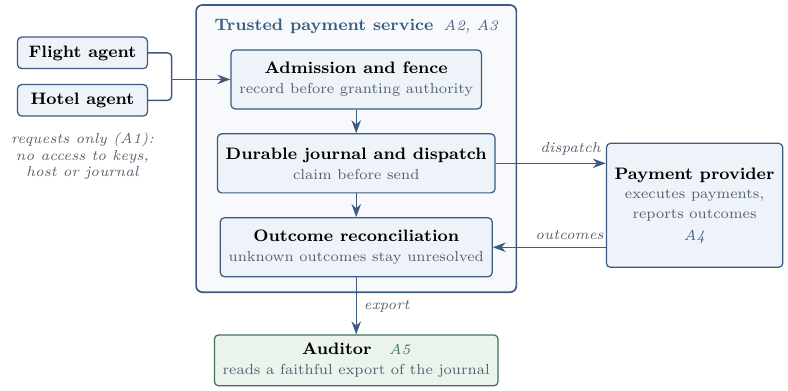}
\caption{Trusted-service architecture. Agents only submit requests; the trusted
service records, grants, fences and reconciles; the provider executes payments
and reports outcomes. A1--A5 mark where each assumption applies.}
\label{fig:architecture}
\end{figure}

\paragraph{Architecture.}
Agents send requests to a single trusted payment service, which holds the
provider credentials, maintains the journal, and talks to the payment provider
(Fig.~\ref{fig:architecture}). Agents are request-generating clients; we do not
evaluate their planning. They may replay, reorder, forge or omit requests, but
cannot access the service's host, keys, journal or administrative controls.

\paragraph{Terminology.}
An \emph{epoch} groups payment operations under a scope fixed before admission.
An \emph{effect} is a payment execution, such as the hotel charge, as distinct
from its request or receipt. An \emph{outstanding right} is authority that can
still produce an effect, including work already sent to the provider. An
\emph{admission fence} blocks new journal entries and grants; it does not cancel
earlier work. To \emph{discharge} a right is to establish that it can no longer
produce an effect, because its effect has completed or its authority has been
irreversibly removed. A timeout, a stopped agent, or a cancellation request
alone is not discharge.

\paragraph{Assumptions.}
The service and the relevant provider contracts are trusted:
\begin{description}[style=sameline,leftmargin=1.2em]
\item[A1 Request-only agents.] Agents cannot access the service's internals.
\item[A2 Complete mediation.] Every effect-capable route in the epoch's scope
passes through admission; identities cannot be reassigned to evade the fence.
\item[A3 Durable ordering.] Recording precedes granting, both serialize with the
fence, and committed state is not rolled back.
\item[A4 Truthful terminality.] Outcomes bind the correct identities, and
discharge irreversibly removes all remaining authority, including in-flight
delivery and provider-side recovery.
\item[A5 Faithful export.] The auditor receives the authoritative journal and
truthful outcomes for the declared scope.
\end{description}
A2 is the critical deployment obligation: a second credential or operator able
to create an attributable effect is an in-scope route that must be mediated.
Signed reports cannot replace mediation, because a signature identifies who
made a report but says nothing about work omitted from it.

\paragraph{Problem statement.}
Four assertions are often used interchangeably. \emph{Admission closed}: no new
records or grants are accepted, though earlier rights may still execute.
\emph{Exposure covered}: reserved funds cover every possible debit, though
outcomes may be unknown. \emph{Execution stopped}: no permitted continuation
adds an effect, though past effects may be unknown. \emph{Exactly closed}: the
admitted population is completely accounted for \emph{and} future effects are
excluded. We seek conditions under which a service can establish exact closure
at the granularity of effect identities.

\paragraph{Following one payment.}
Table~\ref{tab:hotel} follows the hotel payment $h$ through the service.
Approval and command creation belong to the trusted service; an agent can only
ask that an approved command be executed, so a request never becomes authority
merely because an agent submitted it.

\begin{table}[t]
\centering\small
\caption{One hotel payment through the service.}
\label{tab:hotel}
\begin{tabularx}{\linewidth}{@{}>{\raggedright\arraybackslash}p{0.30\linewidth} X@{}}
\toprule
Event & What the service can conclude \\
\midrule
Record $h$, then grant & $h$ is journaled before it has authority; the right is tracked until it is discharged. \\
Claim durably, then send & A retry cannot send $h$ again; the claim alone does not show that payment occurred. \\
Timeout; pause agent; fence & No new records or grants; $h$ may still complete under its earlier right. \\
Provider pays; reply lost & The effect exists although the service does not yet know it. \\
Reconcile $h$ & A truthful terminal outcome accounts for the effect and discharges the right. \\
Check the whole journal & Exact closure needs this for every journaled identity. \\
\bottomrule
\end{tabularx}
\end{table}

\section{Admission Model and Closure Theorem}
\label{sec:theorem}
The admission model has state $(J,I,R,E,F)$: the journal, issued identities,
remaining rights, effect history, and the fence flag. The first four are lists
of identities, initially empty, and $F$ is initially false. $R$ includes
external delivery authority, not merely a client queue; $E$ is ghost history
that the auditor does not observe. Table~\ref{tab:steps} lists the atomic
transitions. A run is a finite sequence of transitions, with no bound on length
or on the identity domain. The model represents one effect per identity.

\begin{table}[t]
\centering\small
\caption{Admission model transitions. Unlisted fields are unchanged.}
\label{tab:steps}
\begin{tabularx}{\linewidth}{@{}l X X@{}}
\toprule
Action & Guard & State change \\
\midrule
Record $i$ & $\neg F$, $i\notin J$ & Prepend $i$ to $J$. \\
Grant $i$ & $\neg F$, $i\in J$, $i\notin I$ & Prepend $i$ to $I$ and $R$. \\
Deliver $i$ & $i\in R$ & Remove $i$ from $R$; prepend it to $E$. \\
Cancel $i$ & --- & Remove $i$ from $R$. \\
Fence & --- & Set $F$ to true. \\
\bottomrule
\end{tabularx}
\end{table}

\begin{theorem}[Fenced coverage and exact closure]
\label{thm:closure}
Let $s$ be reachable from the initial state with $F(s)=\mathrm{true}$, and let
$t$ be reachable from $s$. Then
\begin{align}
J(t)&=J(s),\label{eq:fixed}\\
R(t)\cup E(t)&\subseteq J(s).\label{eq:covered}
\end{align}
If moreover $J(s)\cap R(s)=\varnothing$ and a predicate $\mathrm{reported}$
satisfies $\mathrm{reported}(i)\Leftrightarrow i\in E(s)$ for every $i\in J(s)$,
then
\begin{align}
R(t)&=\varnothing,& E(t)&=E(s),\label{eq:stable}\\
i\in E(t)&\ \Longleftrightarrow\ i\in J(s)\land\mathrm{reported}(i).
\label{eq:exact}
\end{align}
\end{theorem}

\begin{proof}
Every transition preserves $R\cup E\subseteq J$: a grant needs a journal entry,
delivery consumes an existing right, and cancellation adds no effect; the
initial state satisfies it trivially. After the fence, record and grant are
disabled, so the journal is fixed, giving \eqref{eq:fixed}--\eqref{eq:covered}.
Coverage and discharge of every journaled identity imply $R(s)=\varnothing$; no
later transition can add a right or deliver an effect, giving \eqref{eq:stable}.
Truthful reports then identify exactly the executed journal entries, giving
\eqref{eq:exact}.
\end{proof}

\begin{figure}[t]
\centering
\includegraphics[width=\linewidth]{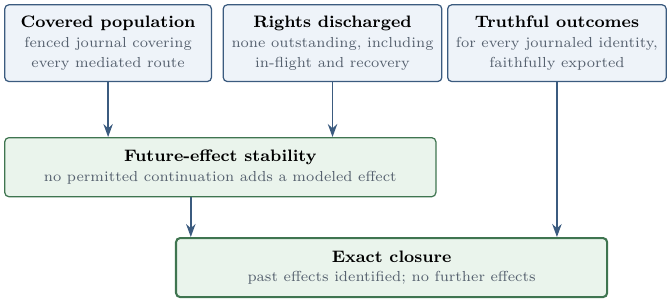}
\caption{Logical dependencies of exact closure. A covered population, fenced
admission and discharge of remaining rights give future-effect stability;
truthful outcomes and a faithful export additionally identify past effects.}
\label{fig:completeness}
\end{figure}

The theorem is mechanized in Lean~4~\cite{lean42021} as four declarations.%
\footnote{\raggedright In \code{Admission.lean}: \code{fenced\_population\_immutable},
\code{post\_fence\_coverage}, \code{drained\_fence\_effects\_stable}, and
\code{reconciled\_fence\_complete}.\par}
For example, the stability component reads:
\begin{verbatim}
theorem drained_fence_effects_stable {s t : State}
    (future : Runs s t) (closed : s.closed = true)
    (drained : s.rights = []) :
    And (t.rights = []) (t.effects = s.effects)
\end{verbatim}
Stability needs drainage but no outcome reports; exact accounting of the past
additionally needs truthful reports. Fig.~\ref{fig:completeness} summarizes
these dependencies.

\paragraph{The hotel example.}
Let $f$ be the completed flight payment and $h$ the hotel payment; after the
fence, $J=\{f,h\}$. Table~\ref{tab:theorem_states} separates the actual effect
history from what the service knows. In the middle row $R$ is already empty, so
no further payment is possible, yet the service still cannot account for the
hotel. Completing a payment and knowing enough to close the books are separate
obligations, which is exactly what the theorem's two parts capture.

\begin{table}[t]
\centering\small
\caption{The hotel example in the admission model. Reconciliation changes the
service's evidence, not the payment history.}
\label{tab:theorem_states}
\begin{tabularx}{\linewidth}{@{}>{\raggedright\arraybackslash}p{0.33\linewidth} c c >{\raggedright\arraybackslash}X@{}}
\toprule
Stage & $R$ & $E$ & Evidence available to the service \\
\midrule
Fence; hotel still pending & $\{h\}$ & $\{f\}$ & Flight paid; hotel unresolved. \\
Hotel pays; reply lost & $\varnothing$ & $\{f,h\}$ & Hotel unresolved despite its payment. \\
Reconciliation & $\varnothing$ & $\{f,h\}$ & Both outcomes known; exact closure. \\
\bottomrule
\end{tabularx}
\end{table}

\paragraph{Counterexecutions.}
Two checked executions show that the premises are needed. In
\code{fence\_without\_drain}, an identity is recorded and granted, admission is
fenced, and the identity is then delivered: a fence alone does not stop
earlier authority, matching the applied-state fence observation of
EffectBound~\cite{effectbound}. In \code{bypass\_breaks\_coverage}, an effect is
added outside the transition relation without changing the journal; this
violates A2, and no journal-based evidence can detect it.

\paragraph{Transfer to implementations.}
A generic result, \code{refinement\_transfers\_coverage}, states what an
implementation must show to inherit coverage: every concrete action simulates
zero or more model steps, every concrete effect and right is represented, the
initial abstraction is initial, and the exported journal equals the abstract
one. This bridge transfers membership coverage; transferring exact accounting
additionally requires an outcome-preserving representation, which we leave to
future work (Sect.~\ref{sec:limitations}).

\section{Provider-Contract Model}
\label{sec:mapping}
Payment providers let a service create a payment object before authorizing
it. The provider-contract model%
\footnote{\raggedright Defined in \code{StripeGate.lean}, named for the Stripe
payment flow that motivated it, in which the service confirms and captures each
payment explicitly (\emph{manual capture}).\par}
extends the admission model with provider objects and terminal identities.
\emph{Allocation} creates an inert object with no payment authority.
\emph{Registration} requires an existing object and an open epoch.
\emph{Dispatch} requires registration, an open epoch, an unused identity, and
no terminal status; it creates a right. \emph{Paid} consumes the right, records
an effect and marks the identity terminal; \emph{cancel} removes the right and
marks it terminal. Lost replies, reads and rejected attempts leave the state
unchanged.

We prove in Lean that every provider transition, and every finite run, maps to
admission-model steps.%
\footnote{\raggedright \code{provider\_step\_refines},
\code{provider\_runs\_refine}, \code{terminal\_manifest\_drains},
\code{terminal\_cut\_stable}.\par}
Consequently an unregistered object never has a right or an effect, a journal
whose entries are all terminal has no remaining rights, and after a fence the
effect history is stable. In practice, a service need not enumerate every
object created at the provider; it must only ensure that no route can activate
an unregistered one, which makes a lost creation response harmless. This
specializes AcquireBound's separation of acquisition from
activation~\cite{acquirebound} to payment objects. Terminality is defined by the
permanent exclusion of future effects, not inferred from a status name; an
operation that can debit more than once needs a finer representation.

\section{Implementation}
\label{sec:impl}
We implemented the discipline as a Python service over SQLite (the
\emph{gate}), with a restricted command interface for agents and a provider
simulator that enforces one effect per identity and irreversible terminality.
\begin{itemize}
\item \emph{Admission.} Registration, grants and the fence are serialized in
SQLite transactions. Agents can only ask for a previously approved command to
be executed, poll status, or request reconciliation.
\item \emph{Claim before send.} A command is durably claimed before it is sent,
so a retry after a crash or lost reply cannot send it again.
\item \emph{Observation integrity.} A report that money was received is accepted
only if a capture command was claimed for that payment. Conflicting terminal
reports, scope mismatches and missing provenance are recorded as durable
conflicts that block closure; a positive receipt is never overwritten by a
later lower report.
\item \emph{Ordered observations.} Each provider retrieval and its journal
update run in one write transaction. A row applied outside it, fetched earlier
or cached, carries a read ticket taken before the read
(Sect.~\ref{sec:concurrent}).
\item \emph{Trace projection.} Recorded service events are translated into
provider-model transitions and checked in Lean (Sect.~\ref{sec:traces}).
\end{itemize}
Table~\ref{tab:assumption_evidence} relates each assumption to the mechanisms
that support it and to what a deployment must still show.

\begin{table}[t]
\centering\small
\caption{How the implementation supports each assumption.}
\label{tab:assumption_evidence}
\begin{tabularx}{\linewidth}{@{}l >{\raggedright\arraybackslash}X >{\raggedright\arraybackslash}X@{}}
\toprule
& Implemented and tested & Remaining deployment obligation \\
\midrule
A1 & Restricted command interface; invalid requests rejected. & Host, key and journal isolation from agents. \\
A2 & Registered identities and approved commands on the tested path. & Mediation of every credential, endpoint, operator and recovery route. \\
A3 & SQLite ordering; claim before send; crash and reopen tests. & Durability under power loss; transport retry discipline. \\
A4 & Simulator terminality; persistent observation conflicts. & Truthful outcomes and discharge at a real provider. \\
A5 & One-transaction snapshot; exported-trace audits. & Complete event capture and faithful export. \\
\bottomrule
\end{tabularx}
\end{table}

\section{Evaluation}
\label{sec:eval}
We address four research questions.
\textbf{RQ1:} Do the proofs check, and do they depend on their premises?
\textbf{RQ2:} Can an auditor detect missing coverage from evidence alone?
\textbf{RQ3:} Do executed service histories conform to the model?
\textbf{RQ4:} Does the service preserve closure under client misbehavior and
concurrency?
Table~\ref{tab:evidence} summarizes the results. All checks run from the public
artifact without provider credentials or network access.

\begin{table}[t]
\centering\small
\caption{Summary of evaluation results. Counts measure different kinds of
checks and should not be summed.}
\label{tab:evidence}
\begin{tabularx}{\linewidth}{@{}>{\raggedright\arraybackslash}p{0.30\linewidth} >{\raggedright\arraybackslash}X@{}}
\toprule
Check & Result \\
\midrule
Lean proofs (RQ1) & 14 exported results checked; 8 premise-removal mutations rejected. \\
Evidence-rule diagnostic (RQ2) & 2,048 records over 64 inputs per rule; 9 tests pass. \\
Service histories (RQ3) & 6 histories, 64 transitions accepted by Lean; 5 closed, 1 undrained; 2 invalid inputs rejected. \\
Generated schedules (RQ4) & 864 cases: 768 resolved, 96 correctly unresolved. \\
Scripted clients (RQ4) & 5 policies, 40 cases; no premature closure or repeat capture. \\
Concurrent workers (RQ4) & 96 cases: 78 resolved, 18 pending, 0 conflicts. \\
Regression tests & 189 run by the reproduction command; 422 in total. \\
\bottomrule
\end{tabularx}
\end{table}

\subsection{RQ1: Formal Checks}
All eight admission results and six provider-contract results check in
Lean~4.19.0 using only the standard axioms \code{propext} and
\code{Quot.sound}; no custom axioms or unfinished proofs are accepted. To test
that the proofs genuinely depend on their premises, the checker removes eight of
them in turn: record-before-grant, record/fence ordering, grant/fence ordering,
delivery authorization, inert creation, registered dispatch, terminal dispatch
exclusion, and source discharge on cancellation. All eight altered proofs are
rejected.

\subsection{RQ2: What Evidence Can Show}
\label{sec:diagnostic}
The diagnostic asks what an auditor can conclude from the evidence it receives:
reports, record sequences and fence claims, but not whether a bypass hid an
effect. A strict rule checks reports, fences and contiguous sequences; a
permissive rule omits the fence and contiguity checks. The generator produces
2,048 labeled records from 64 distinct inputs per rule and keeps the bypass fact
outside the verifier's input. Two histories with identical visible reports,
sequence and fence claim, one of which also contains a bypass payment, receive
the same verdict from any rule restricted to that input. Without a bypass the
strict rule matches the target; with one, its positive verdict can miss an
effect. Coverage must therefore come from mediation (A2), not from inspecting
reports, the problem also identified by Salas~\cite{salas}.

\subsection{RQ3: Service Histories Checked in Lean}
\label{sec:traces}
We ran the Python/SQLite gate through six scenarios against the provider
simulator (Table~\ref{tab:service_traces}) and projected each recorded history
into the provider model: registration and the first durable grant of a command
become registration and dispatch; the service fence becomes the model fence; a
simulator payment commit or cancellation becomes a paid or canceled transition;
claims, replies, reads, reopenings and rejections are stuttering steps. At each
event, the checker compares the projected journal, issued identities and
effects with the gate's rows and the simulator's committed events, and keeps a
right outstanding for every identity without simulator terminality. Lean
accepted all six histories (64 transitions). Five final states satisfy the
closure theorem's premises; the abandoned claim correctly remains undrained.
Two deliberately inconsistent inputs, a payment without a grant and a payment
after terminal cancellation, are rejected.

\begin{table}[t]
\centering\small
\caption{Six executed service histories; all outcomes matched expectations.}
\label{tab:service_traces}
\begin{tabularx}{\linewidth}{@{}p{0.33\linewidth} X@{}}
\toprule
Scenario & Expected and observed outcome \\
\midrule
Cancel without a command & Closed after observing cancellation; no effect. \\
Capture across the fence & Unresolved until the post-fence payment is observed; then closed. \\
Lost reply and reopen & Repeat dispatch rejected; closed after reconciliation. \\
Two payments resolve separately & Unresolved with one pending; closed after its cancellation. \\
Claim abandoned before send & One right remains; closure withheld. \\
Late capture after cancellation & Claim refused after observed cancellation; no provider call; closed. \\
\bottomrule
\end{tabularx}
\end{table}

\subsection{RQ4: Client Behavior and Concurrency}
\label{sec:concurrent}
\paragraph{Generated schedules.}
Starting from one registered, confirmed manual-capture payment, a Python
checker enumerates the 24 orders of grant, fence, dispatch and observation and
the 120 orders with cancellation added. Each order runs with a normal reply, a
reply lost after the payment commits, or a claim abandoned before send, and
with and without reopening both databases after every step:
$(24+120)\times3\times2=864$ cases. An independently written oracle predicts
permitted operations, rights, effects, observations and closure, and a common
suffix checks repeat dispatch, admission after fencing and final reconciliation.
All 864 cases pass: 768 resolve and 96 correctly retain outstanding authority.
The checker's tests reject four broken service variants (missing fence,
repeated claims, premature closure, missing journal entry), and separate
subprocess tests kill the service at 15 points during dispatch.

\paragraph{Scripted clients.}
Agents are modeled as five deterministic request policies driving the actual
command interface. Each receives only an approved command ID: it can neither
register, grant, change amounts, nor supply provider observations. The policies
make one attempt; retry up to twice; poll stored status; reconcile after an
attempt; or submit invalid requests and replay the approved command. Each runs
with a normal reply, a lost reply, indefinitely pending provider work, and
provider cancellation, with and without database reopening: 40 cases
(Table~\ref{tab:agents}).

\begin{table}[t]
\centering\small
\caption{Scripted-client results over eight cases per policy. Every policy
caused exactly eight capture invocations and four paid effects, however many
requests it sent. The last column follows a separate operator reconciliation.}
\label{tab:agents}
\begin{tabularx}{\linewidth}{@{}>{\raggedright\arraybackslash}X r r r@{}}
\toprule
Client policy & Dispatch requests & Client resolves & Operator resolves \\
\midrule
Single attempt & 8 & 0/8 & 6/8 \\
Bounded retry & 20 & 0/8 & 6/8 \\
Snapshot polling & 8 & 0/8 & 6/8 \\
Reconcile after attempt & 8 & 6/8 & 6/8 \\
Invalid and replayed requests & 24 & 6/8 & 6/8 \\
\bottomrule
\end{tabularx}
\end{table}

However aggressively a client retries, each policy causes exactly one capture
per case, and no case closes prematurely. All 40 forged or unsupported
requests sent by the invalid-request policy (five per case) are rejected. The difference between policies is in what they can
\emph{learn}: retries and status polls obtain no fresh provider evidence, so
only reconciling clients resolve cases themselves. Reconciliation establishes
terminal outcomes but correctly cannot discharge genuinely pending work, which
is why 10 cases stay open. This mirrors the theorem's separation between
excluding future effects and knowing past ones.

\paragraph{Concurrent workers.}
We ran separate service workers sharing one journal, each with its own SQLite
connection, in 96 cases (Table~\ref{tab:concurrent}): 2--8 clients contending
over 2--4 payments with normal, lost or pending replies; captures racing
cancellations while the epoch is fenced; and admissions racing the fence in both
transaction orders. Checks at synchronized checkpoints cover journal membership,
claims, per-identity effects, remaining rights and closure.

\begin{table}[t]
\centering\small
\caption{Concurrent-worker outcomes (three repetitions of 32 configurations).
Pending cases correctly retain outstanding authority.}
\label{tab:concurrent}
\begin{tabularx}{\linewidth}{@{}>{\raggedright\arraybackslash}X r r r r@{}}
\toprule
Scenario family & Cases & Resolved & Pending & Conflicts \\
\midrule
Contending dispatch and retries & 54 & 36 & 18 & 0 \\
Capture versus cancellation & 12 & 12 & 0 & 0 \\
Admission versus fencing & 12 & 12 & 0 & 0 \\
Overlapping provider reads & 6 & 6 & 0 & 0 \\
Ticketed unserialized or cached reads & 12 & 12 & 0 & 0 \\
\midrule
Total & 96 & 78 & 18 & 0 \\
\bottomrule
\end{tabularx}
\end{table}

\paragraph{A reconciliation bug found and fixed.}
The first version of this experiment exposed six false conflicts. One
reconciler read a payment as still requiring capture and paused; a second
recorded the terminal outcome; the first then applied its older response. Both
reports were truthful when read, but the gate recorded a conflict and could
never close. The fix holds a write transaction across each provider retrieval
and its journal update, so an older read cannot overtake a newer one. Six
targeted cases pause a pending read, confirm that the lock is held, and let a
competing worker attempt reconciliation; all six close. The cost is that slow
provider reads block journal writers, so adapters must bound timeouts.

\paragraph{Ordering reads without the lock.}
Rows fetched earlier or served from a cache cannot be serialized this way. The
gate therefore keeps a durable per-payment version, incremented whenever an
applied observation changes the record, and every row carries the \emph{read
ticket} (the version) taken before its read was issued. A read issued under
version $v$ was served after every change counted in~$v$. If the version is
still $v$ on application, the row is checked strictly, and a row that looks
older is evidence of provider regression. If the version has advanced, a row
that is only an earlier point of the recorded history (uncertain, with no more
money received) is refused as stale without a conflict; any other difference
is still checked. Twelve cases pause a ticketed read without the lock, directly
or through a cache, while a competing worker records the terminal outcome; all
twelve refuse the older row and close after a retry. With both mechanisms, all
96 cases pass with no conflicts, each identity has at most one capture
invocation and one paid effect, and genuinely contradictory reports still raise
conflicts. Tests reject five broken variants: repeated claims, premature
closure, a missing journal entry, unserialized retrieval, and a gate that
ignores tickets.

\section{Related Work}
\label{sec:related}
\paragraph{Effect closure and mediation.}
EffectBound separates future-use closure, policy-relative effect closure and
quiescence, with Lean checks for a finite-contract reduction~\cite{effectbound};
TrustBound additionally requires exact accounting of past effects from
truthful outcomes. Commit-time authorization requires a complete, current
justification at an atomic sink~\cite{cta}. Salas separates proposal
computation from authoritative admission and explains why a ledger cannot
discover effects that bypass its records~\cite{salas}. AcquireBound quarantines
acquired outputs until activation checks~\cite{acquirebound}, and AID-Guard
conditions recovery on terminal evidence and delivery fencing~\cite{aid}.
Complete mediation~\cite{saltzer1975} underlies our A2.

\paragraph{Payments and accounting.}
Zhu and Wang compose provider frontiers through a shared cut and exact channel
accounting~\cite{quiescence} and treat compensation as a separately authorized
effect~\cite{budget}. Op distinguishes execution rights, provider outcomes,
finality and compensation~\cite{op}. Jiang et al.\ analyze agent payment
protocols (x403, MPP, ACP, AP2) in Tamarin~\cite{protocolanalysis}; those
properties concern individual transactions rather than epoch-wide accounting.
HoldSpec checks a bounded authorization-hold lifecycle~\cite{hold}, and a
closure Internet-Draft distinguishes authenticated receipts from evidence of
complete coverage~\cite{closuredraft}.

\paragraph{Distributed systems foundations.}
Distributed snapshots~\cite{chandy} and Dijkstra--Scholten termination
detection~\cite{dijkstra1980} account for global state and outstanding work; our
rights list abstracts that work. Atomic commit~\cite{gray2006} and
sagas~\cite{sagas1987} coordinate or compensate partial execution;
ARIES~\cite{aries1992} and RIFL~\cite{rifl2015} provide recovery and exactly-once
retries. Our trace checks follow runtime verification, which checks observed
executions against a formal specification~\cite{leucker2009}.
IronFleet~\cite{ironfleet2015} and Verdi~\cite{verdi2015} verify
distributed implementations end to end by refinement, a stronger guarantee than
our trace-level checks.

\section{Limitations and Threats to Validity}
\label{sec:limitations}
\paragraph{Model scope.} The theorem covers one effect per identity on one
trusted payment path, starting from an empty epoch. Partial captures, refunds,
monetary totals and multi-provider composition need a richer model. Booking
success and legal finality are out of scope.
\paragraph{Assumptions.} A1--A5 are deployment obligations, not results.
Table~\ref{tab:assumption_evidence} lists what a real deployment must still
show, most importantly mediation of every effect-capable route (A2) and
truthful, irreversible terminality at the provider (A4). Pending work may never
terminate, in which case closure is correctly withheld.
\paragraph{Implementation.} The Python service is not proved to refine the Lean
models for all executions. The trace checks cover six histories, and the
projection and event capture are trusted. Schedules and concurrent cases are
bounded and do not enumerate all interleavings. Observation ordering still
trusts adapters to take a read ticket before issuing each read and to keep it
with any cached copy, and assumes each provider read reflects every outcome
completed before it was served; stale replicas inside the provider fall under
A4. All experiments use a provider
simulator. A preliminary run against Stripe's test mode
(Appendix~\ref{sec:historical}) predates the service evaluated here and is
reported only as an observation.

\section{Conclusion}
\label{sec:conclusion}
TrustBound separates two obligations that agent-driven payments make easy to
confuse: excluding future payments under earlier authority, and accounting for
payments that already happened. A serialized fence over a record-before-grant
journal fixes the population that must be accounted for; discharging every
outstanding right excludes future effects; truthful outcomes then give exact
closure. The models are checked in Lean, mapped to a provider contract with
inert objects, and connected to an executable service whose histories are
checked against them. Because the discipline depends only on identities,
rights, effects and outcomes, it applies to other autonomous systems whose
effects pass through a trusted component. The most direct next step is an outcome-preserving
refinement proof from the service to the model, which would turn the
trace-level evidence into a guarantee for all executions.

\paragraph{Artifact.}
The Lean models, service, tests and reproduction scripts are available at
\url{https://github.com/tokencraftdata/trustbound}
under the MIT license. Appendix~\ref{app:artifact} gives the reproduction
command.

\paragraph{Use of generative AI.}
OpenAI ChatGPT and Codex, and Anthropic Claude, assisted with literature
exploration, code development and manuscript editing. Reported results come
from recorded tool executions; the authors take responsibility for the paper.

\begingroup
\small
\raggedright

\endgroup

\appendix
\section{Artifact and Reproduction}
\label{app:artifact}
The artifact is the public repository
\url{https://github.com/tokencraftdata/trustbound}. It
contains the Lean sources and checkers, the Python service and tests, reference
traces, and pinned dependencies. With Python~3.11+ and Lean~4.19.0, run from the
repository root:
\begin{verbatim}
python3 -m venv .venv
.venv/bin/python -m pip install -r requirements-reviewer.txt
.venv/bin/python scripts/reproduce_reviewer_artifact.py \
  --lean /absolute/path/to/lean-4.19.0/bin/lean \
  --output results/reviewer-run
\end{verbatim}
Success is exit status zero and \code{checks\_passed: true} in
\code{summary.json}, with the counts in Table~\ref{tab:evidence}. After setup,
no network access or provider credentials are needed. The repository's CI
reruns this command on Ubuntu and macOS for every change, and
\code{docs/REVIEWER\_GUIDE.md} explains every output.

\section{Preliminary Stripe Test-Mode Observation}
\label{sec:historical}
Before building the service evaluated in Sect.~\ref{sec:eval}, we ran a
prototype against Stripe's test mode. The run made nine requests, checked 21
assertions, and captured 200 test cents. It exercised a lost reply, a database
reopen, rejection of a duplicate request, and recovery by re-reading a known
payment from the provider. The prototype differs from the evaluated service in
ways that matter here. Its reconciliation used the same provider key as
dispatch, and its reads were not serialized with journal updates
(cf.\ Sect.~\ref{sec:concurrent}). In addition, the client process ran as
the same operating-system user as the service and could read its key and
journal. The run therefore does not satisfy the request-only boundary of A1 and
does not establish complete admission coverage. It is evidence about the paths
it exercised, not a validation of the provider-contract model or of A1--A5.

A companion local simulation illustrates the limit identified in
Sect.~\ref{sec:diagnostic}. In two runs, one of which adds a payment that
bypasses the service, the service's reads of the payments it knows about are
identical, although 200 synthetic cents are captured in one run and 300 in the
other. The source and saved outputs of both observations are in the
repository. Neither establishes production behavior or conformance with the
Universal Commerce Protocol (UCP) or the Agent Payments Protocol
(AP2).

\section{Supplementary Implementation Details}
\label{app:implementation}
\paragraph{Checkout-bound authorization.}
The service authorizes a payment only from facts extracted from the
authenticated checkout itself, never from facts supplied alongside it. For
example, a signed \$200 checkout paired with separate \$1 facts is rejected.
The service builds checkouts with the official UCP models and verifies
mandates with the AP2 SDK, but this binding rule is our own and does not
establish conformance with either protocol.

\paragraph{Spending allowances.}
A trusted coordinator must atomically claim a task's spending allowance
before any work starts. Retiring an allowance stops further use but does not
free capacity that was claimed, because work may be in flight or funds already
spent. For a \$500 trip, a claimed \$300 allowance therefore stays reserved
after retirement, while an allowance that was never claimed is released. An allowance is a cap, not a money transfer; settlement and
automatic reclamation of claimed capacity are not implemented.

\paragraph{Ledger input validation.}
The spending ledger, which reserves budget atomically for each payment
attempt, rejects negative amounts, booleans, non-integers, and values
outside SQLite's signed 64-bit range before any write; zero is accepted. Tests
cover invalid budgets, capacity creation, identity reuse after rejection,
restart, and integer boundaries. Validation applies to new writes only; rows
already in a database are not checked or repaired.

These modules are covered by the full test suite but not by the
reproduction command of Appendix~\ref{app:artifact}.
Their monetary rules and composed approval-to-dispatch behavior are not proved
by the closure theorem.


\begin{thebibliography}{99}
\bibitem{chandy}
K. Mani Chandy and Leslie Lamport.
Distributed Snapshots: Determining Global States of Distributed Systems.
\emph{ACM Transactions on Computer Systems}, 3(1):63--75, 1985.
\url{https://lamport.azurewebsites.net/pubs/chandy.pdf}.

\bibitem{aid}
Yingzhe Tong, Leyu Dai, and Songhui Guo.
AID-Guard: Stateful Authorization for Delegated Agent Effects.
arXiv:2608.21159v1, 2026.
Sections II--IV, especially III-E.
\url{https://arxiv.org/html/2608.21159v1}.

\bibitem{op}
Raeez Lorgat. Op: Compliance-Carrying Operations.
Unversioned author manuscript, September 2026; accessed 6 November 2026.
The live text may change.
Sections 5.2, 5.8--5.10, especially Proposition 5.7.
\url{https://research.momentum.inc/papers/op.html}.

\bibitem{hold}
Abhishek Sharma.
HoldSpec: A Machine-Checked Model and Conformance Suite for the Payment
Authorization Hold Lifecycle.
arXiv:2610.14419v1, 2026. Sections III-D, V, VIII.
\url{https://arxiv.org/html/2610.14419v1}.
\bibitem{quiescence}
Genliang Zhu and Chu Wang.
Authorization Revocation for Long-Running AI Agents: Root-Scoped Quiescence
under Delegation and Asynchronous Execution.
arXiv:2609.21284v1, 2026. Sections 2.4 and 5.
\url{https://arxiv.org/html/2609.21284v1}.

\bibitem{budget}
Genliang Zhu and Chu Wang.
Fault-Tolerant Budget Conservation in Distributed Multi-Agent Delegation.
arXiv:2610.00349v1, 2026. Sections 4.7--4.8 and 7.6.
\url{https://arxiv.org/html/2610.00349v1}.
\bibitem{effectbound}
Igor Santos-Grueiro.
When Does Authorization End? Effect Closure at Provider Boundaries.
arXiv:2609.02866v1, 2026. Sections II--III, V-C and Appendix C.
\url{https://arxiv.org/html/2609.02866v1}.

\bibitem{salas}
Jesus Salas.
Authority at Commit Time: Reject-and-Rerun Semantics for Governed Agentic Systems.
arXiv:2609.31490v1, 2026. Sections 2.2, 3.2, 6.2 and 7.
\url{https://arxiv.org/html/2609.31490v1}.

\bibitem{acquirebound}
Genliang Zhu and Chu Wang.
Runtime Authorization for Resources Acquired by AI Agents.
arXiv:2609.14744v2, 2026. Sections 4.5, 5.6--5.7; Theorem 14.
\url{https://arxiv.org/html/2609.14744v2}.

\bibitem{protocolanalysis}
Ke Jiang, Mohan Yu, Yuan Chang, Mohit Kumar Jangid, Jianyu Niu,
Cong Wang, and Yinqian Zhang.
A Formal Analysis of Agent Payment Protocols.
arXiv:2609.00060v1, 2026. Table II and Section IV.
\url{https://arxiv.org/html/2609.00060v1}.

\bibitem{cta}
Igor Santos-Grueiro.
Temporary Authority, Permanent Effects: Commit-Time Authorization for LLM Agents.
arXiv:2607.10487v2, 2026. Sections 3--4, especially 4.2--4.3.
\url{https://arxiv.org/pdf/2607.10487v2}.

\bibitem{closuredraft}
D. O. Watts.
Revocation Closure and Authority Epochs for Agent Delegation.
Individual Internet-Draft, draft-watts-oauth-agent-revocation-closure-03,
30 October 2026. Sections 9--11; version, document date and sections verified
on Datatracker, 6 November 2026. Work in progress; not an IETF standard.
\url{https://datatracker.ietf.org/doc/html/draft-watts-oauth-agent-revocation-closure-03}.
\bibitem{saltzer1975}
Jerome H. Saltzer and Michael D. Schroeder.
The Protection of Information in Computer Systems.
\emph{Proceedings of the IEEE}, 63(9):1278--1308, 1975.
\url{https://web.mit.edu/Saltzer/www/publications/protection/}.

\bibitem{dijkstra1980}
Edsger W. Dijkstra and Carel S. Scholten.
Termination Detection for Diffusing Computations.
\emph{Information Processing Letters}, 11(1):1--4, 1980.
\url{https://doi.org/10.1016/0020-0190(80)90021-6}.

\bibitem{gray2006}
Jim Gray and Leslie Lamport.
Consensus on Transaction Commit.
\emph{ACM Transactions on Database Systems}, 31(1):133--160, 2006.
\url{https://www.microsoft.com/en-us/research/publication/consensus-on-transaction-commit/}.

\bibitem{sagas1987}
Hector Garcia-Molina and Kenneth Salem.
Sagas. In \emph{ACM SIGMOD}, pages 249--259, 1987.
\url{https://doi.org/10.1145/38713.38742}.

\bibitem{aries1992}
C. Mohan, Don Haderle, Bruce Lindsay, Hamid Pirahesh, and Peter Schwarz.
ARIES: A Transaction Recovery Method Supporting Fine-Granularity Locking and
Partial Rollbacks Using Write-Ahead Logging.
\emph{ACM Transactions on Database Systems}, 17(1):94--162, 1992.
\url{https://doi.org/10.1145/128765.128770}.

\bibitem{rifl2015}
Collin Lee, Seo Jin Park, Ankita Kejriwal, Satoshi Matsushita, and John Ousterhout.
Implementing Linearizability at Large Scale and Low Latency.
In \emph{ACM SOSP}, 2015.
\url{https://doi.org/10.1145/2815400.2815416}.

\bibitem{ironfleet2015}
Chris Hawblitzel, Jon Howell, Manos Kapritsos, Jacob R. Lorch, Bryan Parno,
Michael L. Roberts, Srinath Setty, and Brian Zill.
IronFleet: Proving Practical Distributed Systems Correct.
In \emph{ACM SOSP}, 2015.
\url{https://doi.org/10.1145/2815400.2815428}.

\bibitem{verdi2015}
James R. Wilcox, Doug Woos, Pavel Panchekha, Zachary Tatlock, Xi Wang,
Michael D. Ernst, and Thomas Anderson.
Verdi: A Framework for Implementing and Formally Verifying Distributed Systems.
In \emph{ACM PLDI}, pages 357--368, 2015.
\url{https://homes.cs.washington.edu/~mernst/pubs/verify-distsystem-pldi2015-abstract.html}.

\bibitem{lean42021}
Leonardo de Moura and Sebastian Ullrich.
The Lean 4 Theorem Prover and Programming Language.
In \emph{Automated Deduction---CADE 28}, LNCS 12699, pages 625--635, 2021.
\url{https://doi.org/10.1007/978-3-030-79876-5_37}.
\bibitem{mastercard2026}
Mastercard.
\href{https://investor.mastercard.com/investor-news/investor-news-details/2026/Mastercard-Advances-Agentic-Commerce-With-New-Trust-and-Intelligence-Services/default.aspx}{Mastercard Advances Agentic Commerce With New Trust and Intelligence Services}.
Company announcement, 30 September 2026; accessed 10 November 2026.

\bibitem{visa2026}
Visa.
\href{https://investor.visa.com/news/news-details/2026/Visa-Releases-First-of-Its-Kind-Industry-Benchmark-for-AI-Shopping-Almost-Half-of-Test-Runs-Fail/default.aspx}{Visa Releases First-of-Its-Kind Industry Benchmark for AI Shopping; Almost Half of Test Runs Fail}.
Company announcement, 5 November 2026; accessed 10 November 2026.


\bibitem{leucker2009}
Martin Leucker and Christian Schallhart.
A Brief Account of Runtime Verification.
\emph{Journal of Logic and Algebraic Programming}, 78(5):293--303, 2009.
\url{https://doi.org/10.1016/j.jlap.2008.08.004}.
\end{thebibliography}
\end{document}